\documentstyle[preprint,eqsecnum,aps]{revtex}

\begin{document}
\draft

\title{Chaos in Schwarzschild Spacetime :\\ The Motion of a Spinning Particle}
\author{{\sc Shingo} SUZUKI\thanks{electronic
mail:696L5186@cfi.waseda.ac.jp} and 
{\sc Kei-ichi} Maeda\thanks{electronic mail:maeda@cfi.waseda.ac.jp}}
\address{Department of Physics, Waseda University,
Shinjuku-ku, Tokyo 169, Japan}
\date{\today}
\maketitle

%%%%%%%%%%%%%%%%%%%%%%%%%%%%%%%%%
%                                  %
%               ABSTRACT        %
%                                  %
%%%%%%%%%%%%%%%%%%%%%%%%%%%%%%%%%
\begin{abstract}
\baselineskip .15in
We study the motion of a spinning test particle in Schwarzschild
spacetime, analyzing the Poincar\'e map and the Lyapunov exponent. 
We find chaotic behavior for a particle with spin higher than
some critical value (e.g. $S_{cr} \sim 0.64
\mu M$ for the total angular momentum
$J=4 \mu M$),  where
$\mu$ and $M$ are the masses of a particle and of a black hole,
respectively.  The inverse of the Lyapunov exponent in
the most chaotic case is about three orbital periods, which
suggests that chaos of a spinning particle may become important in
some relativistic astrophysical phenomena.  

The ``effective potential" analysis enables  us to classify the
particle orbits into four  types as follows. When the total angular
momentum
$J$ is large, some orbits are bounded and the ``effective
potential"s are classified into two types: (B1) one saddle
point (unstable circular orbit) and one minimal point (stable
circular orbit) on the equatorial plane exist for small spin;  and (B2) two 
saddle points bifurcate from the equatorial plane and one minimal point remains on the
equatorial plane for large spin. When $J$ is small, no
bound orbits exist and the potentials are classified into another
two types: (U1) no extremal point is found for small spin; and (U2)  one saddle point appears on the equatorial plane, which is unstable
in the direction perpendicular to the equatorial plane, for large
spin.   The types (B1) and (U1) are the same as those for a
spinless particle, but the potentials (B2) and (U2) are new types caused by spin-orbit coupling.  The chaotic behavior 
is found only in the type (B2) potential. The ``heteroclinic orbit'', which could cause chaos, is also observed in type (B2).
\end{abstract}

\baselineskip .15in

%%%%%%%%%%%%%%%%%%%%%%%%%%%%%%%%%%%%
%                                     %
%           Chapter 1              %
%         Introduction             %
%                                     %
%%%%%%%%%%%%%%%%%%%%%%%%%%%%%%%%%%%%
\section{Introduction}
\label{sec1}\setcounter{equation}{0}
Chaos is now one of the most important ideas used to explain various
non-linear phenomena in nature.  Since the research on the three
body problem by Poincar\'e, many studies about chaos in celestial
mechanics and astrophysics have been done and revealed the
important role of chaos in the Universe\cite{moser},\cite{wisdom}. Although we know many features of  chaos in Newtonian dynamics, we do not
know, so far, so much  about those in general relativity.   If
gravity is strong, e.g., a close binary system or a particle near a black hole, we have to use Einstein's theory of
gravitation. Because the gravitational field  in general relativity 
is non-linear, we may find a new type of chaotic behavior in strong gravitational fields, which do not appear in Newtonian dynamics\cite{misner}-\cite{barrow}. In a previous paper\cite{sota}, we studied a criterion for chaos of a test particle motion around an $N$-black hole system (or an $N$-naked singularity system) and found that a local instability determined by the Riemann curvature tensor provides us a sufficient test for chaos.  We also found that the existence of an unstable circular orbit, which guarantees the existence of a 
homoclinic or heteroclinic orbit, plays a crucial role for
chaos.   However, the relativistic systems  analyzed  so far by
several authors\cite{sota}-\cite{moeckel}, in which chaotic behavior of a test particle is found, are rather unrealistic\cite{sota}-\cite{varvoglis}, except for the perturbed spacetimes of the Schwarzschild black hole solution\cite{bombelli},\cite{moeckel}.  As for the other interesting cases of systems, for example, the $N$-extreme black hole system\cite{contopoulos}-\cite{yustsever} is unstable, the existence of a strong uniform magnetic field around a black hole\cite{karas} is not likely, and naked singularities\cite{sota} may not exist.  We may wonder whether any realistic relativistic system can be chaotic and when chaos may play  an important role in such a relativistic
astrophysical phenomena.

In astrophysics, rotation of a system plays a quite important
role.  The angular momentum or spin may completely change the
evolution of the system. In a dynamical system, rotation or spin is
one of the most important elements and it may sometime cause chaotic behavior.  For example, the H\'enon-Heiles system, which
describes the motion of a star moving in the potential of a
rotating galaxy, is chaotic\cite{henon}. We also know that some spin-orbit interaction induces chaos in Newtonian gravity.  This may also be true in a relativistic system such as the evolution of a
binary system. The motion of coalescing binary systems of neutron
stars and/or black holes is very important to  study because they
are promising sources of gravitational waves, which we are
planning to  detect by large-scale laser interferometric
gravitational observatories, such as US LIGO\cite{abromovici}. If we will detect the signal
of gravitational waves emitted from these systems and compare it
with theoretical templates, we may be able to determine a variety
of astrophysical parameters of the sources such as their direction,
distance, masses, spin, and so on\cite{cutler}.
In order to extract exact information about such sources from the observed signal we need the exact theoretical templates of the gravitational waveforms. To  make such templates, it is very
important to know the exact motion of sources. Hence, the equations of motion in the post-Newtonian expansion in terms of a small parameter
$\epsilon\approx (v/c)^2 \sim GM/r$ have been studied  by many authors\cite{damour}.
Those can be written schematically as
\begin{equation}
\frac{d^2\mbox{\boldmath$x$}}{dt^2}=\mbox{\boldmath$a$}_{N}
+\mbox{\boldmath$a$}_{PN}^{(1)}+\mbox{\boldmath$a$}_{SO}^{(3/2)}
+\mbox{\boldmath$a$}_{2PN}^{(2)}+\mbox{\boldmath$a$}_{SS}^{(2)}
+\mbox{\boldmath$a$}_{RR}^{(5/2)}+O(\mbox{\boldmath$a$}^{(3)}),
\label{1.1}
\end{equation}
where the subscripts $N, PN, SO, 2PN, SS$ and $RR$ denote 
Newtonian, post-Newtonian, spin-orbit coupling,
2nd post Newtonian, spin-spin coupling, and radiation reaction
terms, respectively\cite{will}; and the superscript corresponds to the order of
expansion in $\epsilon$. To make sufficient templates, we may need at least the 3rd-order post Newtonian contribution to obtain  the S/N ratio required from the observation\cite{3rd1}-\cite{3rd3}.  But this is still under
investigation in the world. 

The spin effect is also important. The spin terms in Eq.(\ref{1.1}), $\mbox{\boldmath$a$}_{SO}$ and $\mbox{\boldmath$a$}_{SS}$, induce a precession of the orbital plane through the spin-orbit or spin-spin coupling, resulting in modulation of the gravitational waveforms
\cite{kidder},\cite{apostlatos}. In \cite{apostlatos}, it is also shown that the orbital plane may behave  very strangely  due to the spin effects. We  cannot verify whether or not any chaotic behavior occurs  in their system. But from  the studies on spin effects in
Newtonian dynamics, we know that a spin effect can make a motion
chaotic. We then expect that a relativistic system such as  a coalescing binary pulsar may also show the similar non-linear
phenomena. The gravitational waveform from the
system with chaotic motion will be different from a system with a regular motion, for example, a regular 
precession of the orbital plane as shown in \cite{kidder}. 
The chaos might be too strong to make a complete template of
gravitational waves, or rather it might give us new information about
astrophysical parameters from a time series of the observed waveforms.  We will discuss this problem for a
coalescing binary system with highly spinning bodies elsewhere.

Thus we believe that a study about spin effects on the orbital
evolution of a relativistic system and its gravitational waveform
is very important from the viewpoint of observations as well as of
academic interest. In this paper, to clarify the spin effect on
the orbital motion, especially the spin-orbit interaction, we study the motion of a spinning test particle around a
Schwarzschild black hole. So far, studies about a spinning
test particle in relativistic spacetime have been done by many
authors since the basic equations were derived by
Papapetrou\cite{papa} and reformulated by Dixon\cite{dixon}.
Corinaldesi and Papapetrou already discussed a spinning
test particle in Schwarzschild spacetime\cite{papa2}. But, apart
from the supplementary condition, from which they adopted a
different equation from the present standard one,  they presented
the basic equations and discussed some terms with
physical interpretations. They did not analyze the orbits in
detail from a viewpoint of the dynamical system.  Kerr or Kerr-Newman spacetime was also analyzed
by several authors\cite{Rasband}-\cite{Rudiger}. In
\cite{Rasband},\cite{Tod} and \cite{hojman},  the effective potential of the spinning particle is given and the spin effects on the
binding energy are discussed. In \cite{mino} and \cite{mino2},  the gravitational waves produced by a spinning
particle falling into a Kerr black hole or moving circularly
around it is discussed and the energy emission rate from those
systems is calculated. But in those
papers they discussed only the case of the orbit in  the
equatorial plane or on the symmetric axis of the black hole.
Since we  are interested in chaotic motion induced by spin-orbit coupling here, we have to discuss the most generic
situation, i.e., the orbital motion off the equatorial plane.

 This paper is organized as follows. In section 2 we shall briefly review
the basic equations, i.e., the equations of
motion for a spinning test particle in relativistic spacetime, a supplementary condition and
some constants of motion.  We specify the
background spacetime to be a Schwarzschild black hole, then we write down those equations and introduce a sort of  ``effective potential", which enables us to classify the particle behavior. In section 3, performing
numerical integrations, we show that chaos occurs
for a highly spinning test particle. Summary and some
remarks follow in section 4.

  Throughout this paper we use units $c=G=1$.  We define the signature of the metric as $(-,+,+,+)$.
%%%%%%%%%%%%%%%%%%%%%%%%%%%%%%%%%%%
%                                    %
%              Chapter 2          %
%             Basic Equations     %
%                                    %
%%%%%%%%%%%%%%%%%%%%%%%%%%%%%%%%%%%
%%%%%%%%%%%%%%%%%%%%%%%%%%%%%%%%%%%
%                                    %
%              Chapter 2-1        %
%             Basic Equations     %
%                                    %
%%%%%%%%%%%%%%%%%%%%%%%%%%%%%%%%%%%
\section{Basic Equations for a Spinning Test Particle}
\label{sec2}\setcounter{equation}{0}
\subsection{Pole-dipole Approximation}
The equations of motion of a spinning test particle in a
relativistic spacetime were first derived by Papapetrou\cite{papa}
and then reformulated by Dixon\cite{dixon}.
Those are a set of equations:
\begin{eqnarray}
\frac{dx^{\mu}}{d\tau}&=&v^{\mu},
\label{eqn:xdot}
\\
\frac{Dp^{\mu}}{D\tau}&=&
-\frac{1}{2}R^{\mu}_{~\nu\rho\sigma}v^{\nu}S^{\rho\sigma},
\label{eqn:pdot}
\\
\frac{DS^{\mu\nu}}{D\tau}&=&p^{\mu}v^{\nu}-p^{\nu}v^{\mu},
\label{eqn:sdot}
\end{eqnarray}
where $\tau, v^{\mu}, p^{\mu}$ and $S^{\mu\nu}$ are an affine
parameter of the orbit, the 4-velocity of the particle, the
momentum, and 
the spin tensor, respectively.   $\tau$ is chosen as the proper time
of the particle in this paper, then $ v^{\mu} v_{\mu} = -1$. The multipole moments of the particle higher than mass monopole and
spin dipole are ignored. It is called the pole-dipole
approximation. 

We need  a supplementary condition which gives a relation between
$v^{\mu}$ and  $p^{\mu}$, because $p^{\mu}$ is no longer parallel
to $v^{\mu}$ in the present case. The r.h.s. of
Eq.(\ref{eqn:pdot}) denotes a spin-orbit coupling through a strong
gravitational  field. We adopt the following condition\cite{dixon},
\begin{equation}
p_{\mu}S^{\mu\nu}=0.
\label{eqn:ps0}
\end{equation}
This condition is related to how to choose the center of mass in
an extended body, and this choice gives a consistent
condition\cite{Beiglbock}.  Using (\ref{eqn:ps0}) we can write down the
relation between $v^{\mu}$ and $p^{\mu}$ explicitly, that is,
\begin{equation}
v^{\mu}=N\left[u^{\mu}
+\frac{1}{2\mu^2\Delta}S^{\mu\nu}u^{\lambda} 
R_{\nu\lambda\rho\sigma}
S^{\rho\sigma}\right],
\label{eqn:p-v}
\end{equation}
where 
\begin{equation}
\Delta=1+\frac{1}{4\mu^2}R_{\alpha\beta\gamma\delta}
S^{\alpha\beta}S^{\gamma\delta},
\end{equation}
and
\begin{equation}
N=\left[ 1- \frac{1}{4\Delta^2\mu^2} S_{\mu \nu} u_{\lambda}
S_{\rho\sigma}R^{\nu\lambda\rho\sigma}S^{\mu \alpha} u^{\beta}
S^{\gamma\delta}R_{\alpha\beta\gamma\delta}
\right]^{-1/2}
\end{equation}
 is a normalization constant fixed by 
$v_{\nu}v^{\nu}=-1$.
 $u^{\nu} \equiv p^{\nu}/\mu$ is a unit vector parallel to the
momentum 
$p^{\nu}$, where the mass of the particle $\mu$ is defined by 
\begin{equation}
\mu^2=-p_{\nu}p^{\nu}.
\label{eqn:mass}
\end{equation}
This system has several conserved quantities.
Regardless of the symmetry of the background spacetime, it is 
easy to show that $\mu$  and the
magnitude of spin $S$, defined by  
\begin{equation}
S^2\equiv\frac{1}{2}S_{\mu\nu}S^{\mu\nu},
\label{eqn:spin}
\end{equation}
are  constants of motion\cite{Wald2}.
If a geometry possesses some symmetry described by a  Killing 
vector $\xi^{\mu}$ associated with the symmetry, we can show
that
\begin{equation}
C\equiv\xi^{\mu}p_{\mu}-\frac{1}{2}\xi_{\mu ;\nu}S^{\mu\nu}
\label{eqn:killing}
\end{equation}
is also conserved\cite{dixon}.

%%%%%%%%%%%%%%%%%%%%%%%%%%%%%%%%%%%
%                                    %
%              Chapter 2-2        %
%       Schwarzschild spacetime   %
%                                    %
%%%%%%%%%%%%%%%%%%%%%%%%%%%%%%%%%%%
\subsection{Spinning Particle in Schwarzschild Spacetime}
As for the background spacetime, we assume a  Schwarzschild
spacetime, i.e.,
\begin{equation}
ds^2=-f(r)dt^2
+f(r)^{-1}dr^2
+r^2(d\theta^2+\sin^2\theta d\phi^2),
\end{equation}
where 
\begin{equation}f(r) =1-\frac{2M}{r}
\end{equation} 
with $M$
being the mass of the black hole. Because the spacetime is static
and spherically symmetric, there are  two Killing vector fields,
$\xi_{(t)}^{\mu}$ and
$\xi_{(\phi)}^{\mu}$. From (\ref{eqn:killing}), we find the
constants of motion related with those Killing vectors as 
\begin{eqnarray}
E & \equiv & - C_{(t)}=-p_t-\frac{M}{r^2}S^{tr},
\label{eqn:energy}
\\
J_{z} & \equiv & C_{(\phi)}=p_{\phi}-r (S^{\phi
r}-rS^{\theta\phi}\cot\theta )\sin^2\theta.
\label{eqn:jz}
\end{eqnarray}
$E$ and $J_z$ are interpreted as the energy of the particle and
the $z$ component of the total angular momentum,  respectively.
Because the spacetime is spherically symmetric, the  $x$ and $y$
components of the total angular momentum are also
conserved. Then we have two additional constants of motion as
\begin{eqnarray}
J_{x}&=&-p_{\theta}\sin\phi -p_{\phi}\cot\theta\cos\phi 
\nonumber \\ &&+r^2S^{\theta\phi}\sin^2\theta\sin\phi
+rS^{\phi r}\sin\theta\cos\theta\sin\phi  +rS^{r\theta}\cos\phi,
\label{eqn:jx}
\end{eqnarray}
\begin{eqnarray}
J_{y}&=&p_{\theta}\cos\phi -p_{\phi}\cot\theta\sin\phi 
\nonumber\\
 &&+r^2S^{\theta\phi}\sin^2\theta\cos\phi
+rS^{\phi r}\sin\theta\cos\theta\cos\phi  -rS^{r\theta}\sin\phi.
\label{eqn:jy}
\end{eqnarray}
Because the background is spherically symmetric,  without loss of
generality we can choose the $z$ axis in the direction of total angular momentum
as 
\begin{equation}
(J_{x},J_{y},J_{z})=(0,0,J),
\label{eqn:total}
\end{equation}
where $J>0$.
Three constraint equations  (\ref{eqn:jz})$\sim$(\ref{eqn:jy}) with Eq. (\ref{eqn:total}) are reduced  to 
\begin{eqnarray}
S^{\theta\phi}&=&\frac{J}{r^2}\cot\theta,
\label{eqn:jx2}
\\
S^{r\theta}&=&-\frac{p_{\theta}}{r},
\label{eqn:jy2}
\\
S^{\phi r}&=&\frac{1}{r} \left(-J+\frac{p_\phi}{\sin^2
\theta}\right).
\label{eqn:jz2}
\end{eqnarray}

$S^{ti}\;(i=r,\theta,\phi)$ are fixed from Eq.
(\ref{eqn:ps0}) with Eqs. (\ref{eqn:jx2})$\sim$(\ref{eqn:jz2}) as
\begin{eqnarray}
S^{t r }&=&
-\frac{1}{r p_t} \left(p_\theta^2 +
\frac{p_\phi^2}{\sin^2\theta} -J p_\phi \right),
\label{eqn:str} \\ 
S^{t \theta }&=&
\frac{1}{r p_t} \left(p_rp_\theta + \frac{J p_\phi}{r} \cot
\theta\right),
\label{eqn:stth} \\
 S^{t \phi}&=&
-\frac{1}{r p_t} \left( J p_r - \frac{p_rp_\phi}{\sin^2
\theta} + \frac{J p_\theta}{r} \cot \theta\right).
\label{eqn:stph}
\end{eqnarray}

The energy conservation Eq.(\ref{eqn:energy}) is now
\begin{equation}
E=-p_t + \frac{1}{p_t r^3}\left(p_\theta^2
+\frac{p_\phi^2}{\sin^2\theta} -Jp_\phi \right).
\label{eqn:energy2}
\end{equation}

The procedure to give  initial conditions for calculating the
orbital evolution of the particle is as follows. First, we give
constants of motion $S$, $J$ and $E$ with which the motion of the
particle is bounded to a compact region of the spacetime. We will
discuss how to do it  in detail in the next subsection.  We set the
particle on the equatorial plane $\theta = \pi/2$ with
$r=r_{0}$ and $\phi=0$.  Next, we give the spatial components of
the spin tensor $\mbox{\boldmath$S$}=(S^{\theta\phi}, S^{\phi
r}, S^{r\theta})$.  $S^{\theta\phi}$ vanishes
at $\theta = \pi/2$ from (\ref{eqn:jx2}), which means that the
initial spin is perpendicular to the radial direction. The
last parameter we need to choose initially  is,
\begin{equation}
\alpha\equiv\tan^{-1}\frac{S^{r\theta}}{S^{\phi r}},
\label{eqn:angle}
\end{equation}
which  determines the direction of the spin. 
Note that 
$\alpha=0$ and $\pi$ denote the spin  anti-parallel and
parallel to the positive $z$-direction, respectively. In this case
the orbit of the particle is always restricted to the equatorial
plane and chaos never occurs. We assume
$\pi/2\leq\alpha\leq3\pi /2$, which corresponds to
the case that  the $z$ component of spin points to the same
direction as that of the total angular momentum.  Otherwise, the
particle  cannot get into any relativistic region and chaos
never occurs, as we will see later. 

From Eq. (\ref{eqn:spin})  with Eqs. (\ref{eqn:jx2})$\sim $ (\ref{eqn:angle}), we find that
\begin{equation}
S^{\phi r}= -\frac{p_tS}{\sqrt{J^2 \sin^2\alpha + \mu^2 r_0^2}}\cos\alpha,\hskip 1cm 
S^{r \theta}= -\frac{p_tS}{\sqrt{J^2 \sin^2\alpha + \mu^2 r_0^2}} \sin
\alpha.
\label{eqn:spin3}
\end{equation}
Inserting Eqs. (\ref{eqn:jy2}) and (\ref{eqn:jz2}) with these equations into Eq. (\ref{eqn:energy2}) (at $\theta=\pi/2$), we get the quadratic equation for $u_t$.
Solving it and inserting the result into Eq. (\ref{eqn:spin3}), we find the initial values for $p_t, \mbox{\boldmath$S$}$. $p_{\theta}, p_{\phi}$ and $S^{ti}$ are determined from Eqs. (\ref{eqn:jy2}) $\sim$ (\ref{eqn:stph})
and $p^{r}$ is given by use of Eq. (\ref{eqn:mass}).

%%%%%%%%%%%%%%%%%%%%%%%%%%%%%%%%%%%%%%%%%%%%%%%%%%%%%%%%%%%
%                                                         %
%        Section 2-3      (Effective Potential)           %
%                                                         %
%   Fig.1 : Effective Potential on Equatorial Plane       %
%   Fig.2 : Four Effective Potentials                     %
%               a:Single S.P. b:Double S.P.               %
%               c:Normal Unbound  d:Strange unbound       %
%   Fig.3 : Typical orbit in type (U2) potential          %    
%   Fig.4:  J-S plane                                     %
%                                                         %
%%%%%%%%%%%%%%%%%%%%%%%%%%%%%%%%%%%%%%%%%%%%%%%%%%%%%%%%%%%
\subsection{Contour of Zero Meridian Momentum as an ``Effective
Potential"} 
Since we are interested in chaotic behavior of a particle orbit, we analyze
a test  particle which does not escape to  infinity and does not
fall into a black hole.  Therefore we have to choose appropriate
parameters of the particle, $E$, $J$ and $S$. If we have an
effective potential, such a choice is easy.  For example, for a
spinless particle traveling around a Schwarzschild black hole with
the orbital angular momentum
$L$, whose motion is restricted to a plane
(e.g. the equatorial plane $\theta=\pi /2)$, the effective potential of the particle
is given as 
\begin{equation}
V^2(r;L)=\left(1-\frac{2M}{r}\right)\left(\mu^2
+\frac{L^2}{r^2}\right).
\label{eqn:schwapot}
\end{equation}
Since the region where the particle with the energy $E$ can move
is given by $V^2(r) < E^2$, it is easy to choose  the energy $E$
such that the particle will move in a compact region. 

 In the case of a spinning particle, because the spin-orbit
coupling (\ref{eqn:pdot}) is not a potential force and an
additional dynamical variable, i.e., the direction of the spin, exists, we cannot find any effective potential in the 2-dimensional $r$-$\theta$ plane. To know the region
where the particle can move, however,  we do not  need to find an
effective potential itself.  Rather we need to know only the boundary of such a region, i.e., a curve in the $r$-$\theta$ plane
where  both
$p^r$ and
$p^\theta$ vanish.  From (\ref{eqn:jy2}),  $S^{r\theta}$ also
vanishes there, which  means that the spin lies in the
meridian plane and there remains no freedom of the spin
direction ($\alpha=0$ or $\pi$).  From (\ref{eqn:mass}) with
$p^{r}=p^{\theta}=0$, we can set  
\begin{eqnarray}
p_t &=& - \mu f^{1/2} \cosh X, \nonumber \\
p_\phi&=& \mu r \sin \theta  \sinh X , 
\end{eqnarray}
where $X(r, \theta)$ is an unknown function. Inserting this into
Eq. (\ref{eqn:spin}) with Eqs.  (\ref{eqn:jx2}) $\sim$  
(\ref{eqn:jz2}), we find the
equation for $X$ as
\begin{equation}
(\mu^2r^2 -S^2 f) \sinh^2 X - 2 \mu J r \sin \theta \sinh X +
(J^2 -S^2) f + \frac{2M}{r}J^2 \sin^2 \theta =0. 
\end{equation}
Finally, from Eq.(\ref{eqn:energy}), 
 we obtain the equations for such a curve as 
\begin{equation}
E=V_{(\pm)}(r,\theta;J,S),
\label{eqn:effpot}
\end{equation}
where 
\begin{equation}
V_{(\pm)}(r,\theta;J,S)=
\mu \left[f^{\frac{1}{2}}\cosh X_{(\pm)}
+ \frac{M\sinh X_{(\pm)}}{f^{1/2}r \cosh X_{(\pm)}}
\left(\frac{J \sin \theta}{\mu r} - \sinh X_{(\pm)}
\right) \right], 
\label{eqn:effectivepotential}
\end{equation}
and 
\begin{equation}
\sinh X_{(\pm)} \equiv \frac{\mu Jr\sin\theta}{\mu^2r^2-S^2f}
\pm\left[\frac{\mu^2J^2r^2\sin^2\theta}{(\mu^2r^2-S^2f)^2}
-\frac{\left((J^2-S^2)f+\frac{2M}{r}J^2\sin^2\theta
\right)}{\mu^2r^2-S^2f}\right]^{\frac{1}{2}}.
\end{equation}
 The subscript $(\pm)$ corresponds to
the direction of the spin,  i.e., if the $z$ component of the
spin is the same direction as the total angular momentum
($S^{\phi r}<0$, i.e., $\pi/2 < \alpha < 3 \pi /2$), we take 
$V_{(-)}$, while if it is the opposite ($S^{\phi r}>0$, i.e.,
$-\pi/2 < \alpha < \pi /2$),   $V_{(+)}$ should be applied.
Note that $S^{\phi r}$ never changes its signature during the
evolution. Imposing $S=0$ and $\theta=\pi /2$,
$V_{(\pm)}$ is reduced to
the conventional effective potential for a
spinless particle in a Schwarzschild black
hole, i.e., Eq. (\ref{eqn:schwapot}). 
The derivation also shows that the particle with energy $E$ 
can move in the region of the $r$-$\theta$
plane such that  $V^2_{(\pm)}(r,\theta;J,S) < E^2$.  Then we shall call 
$V(r,\theta;J,S)$  the ``effective potential" of a spinning
particle in Schwarzschild spacetime. 

The typical shape of the potential on the equatorial plane is shown in Fig.1. From this figure,  we can see that the particle in the potential
$V_{(-)}$ can move closer to the event horizon $(r=2M)$ than
that in the potential $V_{(+)}$. This is because  the spin-orbit
interaction is a repulsive force if the spin is parallel to the
orbital angular momentum and it will balance with  the
gravity, so that a stable periodic orbit closer to the
event horizon becomes possible  
without falling into the black hole. If the direction of the spin
is opposite, we find the reverse.  
Such a spin-orbit interaction is induced through the
gravitational interaction as the r.h.s of Eq. (\ref{eqn:pdot}),
which breaks the integrability of the particle motion.
Since we are interested in chaotic behavior, a large
spin-orbit interaction may be more interesting.
Hence, in this paper,  we study only particle motion 
in a strong gravitational field, i.e., in the potential $V_{(-)}$.
In what follows, we drop the subscript $(-)$. 

Once $J$ and $S$ are given, we can depict a contour map of
$V(r,\theta;J,S)$, with which we find an allowed region where a particle with $E$ can move. We find that  the ``effective
potential" $V$ is classified into four types depending on $J$
and $S$. These are given in Fig.2. In Fig.2a, the potential
has one saddle and one minimal point on the equatorial
plane, which  appears when $S$
is small compared with $J$. Since the spin effect is small,
the shape of the potential is similar to that for a spinless
particle. We call this potential type (B1). The orbit in
this case never becomes  chaotic.  Fig.2b shows the
potential which has two saddle points off the 
equatorial plane and one minimal point on the equatorial plane.
The saddle points are located symmetrically on opposite sides of the equatorial plane. 
We call this potential type (B2). For fixed
$J$, as $S$ gets larger, type (B1) changes into type (B2).
This happens because  a repulsive spin-orbit interaction is
angle dependent.  The ``effective potential"s shown in both Figs.2c and 2d have no bound region, although the latter case has a saddle point on the
equatorial plane. The particle will eventually fall into a black
hole. We call those potentials type (U1) and (U2),  respectively.
These potentials appear when $J$ is small enough, i.e., the
centrifugal force is too small to balance with the gravity.
 The type (U1) potential is similar to that for a spinless
particle. In the case of type (U2), however,  the situation is slightly different. Because the saddle point marked by a cross in Fig.2d is minimal
in the $r$ direction and maximal in  the $\theta$ direction,  a
particle will gradually depart from the equatorial plane and fall
into a black hole.   There is a potential barrier on the equatorial
plane from the repulsive spin-orbit interaction.

We show in Fig.3  what values of  $J$ and $S$ belong to which
types. Because types (B2) and (U2) never appear in the
case of a spinless particle,  we conclude that such strange behavior in those potentials is induced by a spin effect.  As we will show
later, we find that it is only for the
type (B2) that the particle motion can be  chaotic.
  Note that the value of $S$ at the
bottom end of the (U2) region is slightly smaller than $1\mu M$, i.e., at $S\sim 0.987\mu M$. 
Hence the type (U2) still appears with physically meaningful
values of $S$ and $J$, as we will discuss soon. 

%%%%%%%%%%%%%%%%%%%%%%%%%%%%%%%%%%%
%                                    %
%          Chapter 3              %
%         Numerical Results       %
%                                    %
%%%%%%%%%%%%%%%%%%%%%%%%%%%%%%%%%%%
\section{Numerical Results}  
\label{sec3}\setcounter{equation}{0}
%%%%%%%%%%%%%%%%%%%%%%%%%%%%%%%%%%%%%%%%%%%%%%%%%%%
%                                                 %
%      Section 3-1 (Poincar\'e Map)               %
%                                                 %
%      Fig.5 : Poincar\'e Map in (B2)            %
%          a:S=0.4, b:S=0.6                       %
%  c:S=0.8, d:S=1.0, e:S=1.2 f:S=1.4              %
%      Fig.6 : Orbits for Fig.5d                  %
%      Fig.7 : Lyapunov Exp.                      %
%%%%%%%%%%%%%%%%%%%%%%%%%%%%%%%%%%%%%%%%%%%%%%%%%%%
\subsection{Chaos in Schwarzschild black hole}
Using the
Bulirsch-Stoer method \cite{BS}, we integrate the equations of
motion numerically  for various values of parameters, $E$, $J$ and
$S$ and various initial configurations
 of $r_{0}$ and
$\alpha$. As for the dynamical variables, we solve all
components of the position vector $x^\mu$, of the momentum
$p^\mu$, and of the spin tensor $S^{\mu\nu}$, and use the
constraint equations
(\ref{eqn:ps0}),(\ref{eqn:mass}),(\ref{eqn:spin}),
(\ref{eqn:energy}) $\sim$
(\ref{eqn:jy}) to check the accuracy of our numerical
integration. 
After $10^{3}$ orbital periods, which is the end of our
numerical calculation,  the relative errors are smaller than 
$10^{-11}$ for each constraint. The
methods with which we analyze the chaotic behavior are the
Poincar\'e map and the Lyapunov exponent. To make the Poincar\'e
map, we adopt the equatorial plane ($\theta = \pi /2$) as a
Poincar\'e section and plot the point $(r,v^{r})$ when the particle
crosses the Poincar\'e section with $v^{\theta}<0$.
 If the motion  is not chaotic, the plotted points form
a closed curve in the 2-dimensional $r$-$v^{r}$ plane, because a regular orbit will move on
a torus in the phase space and the curve is a cross section of
the torus. If the orbit is chaotic, some of those tori will be broken
and the Poincar\'e map does not consist of a set of closed curves
but the points will be distributed randomly in the allowed region. From
the distribution of the points in Poincar\'e map, we can judge
whether  or not the motion is chaotic.
The Lyapunov exponent is another method to judge the
occurrence of chaos and it also gives a naive estimation of  the strength
of chaos quantitatively.
 This denotes how fast the close orbits in the
phase space will diverge in future. 
The Lyapunov exponent
$\lambda$ is defined by
\begin{equation}
\lambda\equiv\lim_{t\rightarrow\infty}\frac{1}{t}
\ln\left|\frac{d(t)}{d(0)}\right| , 
\label{def:Lyapunov}
  \end{equation}
  where $d(t)$ is the distance  at time $t$ between  two
neighboring  points  in the phase space.  If
the orbit is chaotic, then $\lambda$ will converge to some
positive value, which means that
the distance will diverge exponentially
with a typical time scale 
$\tau_{\lambda} \equiv \lambda^{-1}$. For a particle moving along a geodesic, we can
calculate the Lyapunov exponent by integration of the equation of
geodesic deviation; 
\begin{equation}
\frac{D^2n^{\mu}}{D\tau^2}=
-R^{\mu}_{~\nu\rho\sigma}
v^\nu n^{\rho}v^\sigma , 
\label{eqn:geodev}
\end{equation}
where $n^{\mu}$ is a deviation vector. In the present case,
however,  because a spinning test particle does not move along a
geodesic as discussed in section 2, we need another way to
estimate the Lyapunov exponent. Here, we have adopted the method
developed by Sano and Sawada\cite{SS} to
estimate the Lyapunov exponent. 
We prepare a time series of
$\rho(t)$, $z(t)$, $v^{\rho}(t)$ and $v^{z}(t)$, where $\rho$ and
$z$ are defined by $\rho\equiv r\sin\theta$ and $z\equiv
r\cos\theta$, and calculate the Lyapunov exponent from such a set
of data. The way to calculate the Lyapunov exponent is as follows.
Let $\{\mbox{\boldmath$x$}_{i}|i=0,1,2,\cdots\}$ denote a set of points on some orbit in the 4-dimensional phase space, i.e., $\mbox{\boldmath$x$}_{i}=(\rho(i\Delta t),z(i\Delta t),v^{\rho}(i\Delta t),v^{z}(i\Delta t))$ where $\Delta t$ is a small time interval. A set of ``deviation vectors'' $\{\mbox{\boldmath$y$}_{i}\}$ at a point $\mbox{\boldmath$x$}_{i}$ is defined by 
\begin{equation}
\{\mbox{\boldmath$y$}_{i}\}=\{\mbox{\boldmath$x$}_{k_i}-\mbox{\boldmath$x$}_{i}\;|\;\;0<\|\mbox{\boldmath$x$}_{k_i}-\mbox{\boldmath$x$}_{i}\|<\epsilon\},
\end{equation}
where $\epsilon$ is a small constant chosen appropriately. After the evolution of a time interval $m\Delta t$, the orbital point $\mbox{\boldmath$x$}_{i}$ will proceed to $\mbox{\boldmath$x$}_{i+m}$ and neighboring points $\mbox{\boldmath$x$}_{k_i}$ to $\mbox{\boldmath$x$}_{k_i+m}$. The deviation vectors $\mbox{\boldmath$y$}_{i}$ are thereby mapped to
\begin{equation}
\mbox{\boldmath$y$}^{(m)}_{i}=\mbox{\boldmath$x$}_{k_i+m}-\mbox{\boldmath$x$}_{i+m},
\end{equation}
which transformation is described by a matrix $A^{(m)}_{i}$.
Then the Lyapunov exponent $\lambda$ is calculated by
\begin{equation}
\lambda=\lim_{n\rightarrow\infty}\frac{1}{nm\Delta t}\sum_{i=1}^{n}\ln\|A^{(m)}_{i}\mbox{\boldmath$n$}^{i}\|,
\end{equation}
where $\{\mbox{\boldmath$n$}^{j}\}$ is a set of arbitrary unit vectors. If the system is chaotic, $\lambda$ will be positive and will not depend on choice of $\Delta t$, $\epsilon$, $m$ and $\{\mbox{\boldmath$n$}^{i}\}$.

Fig.4 shows the Poincar\'e maps for the total angular momentum 
$J=4 \mu M$ and for several values of the spin  $S$
= $0.4 \sim 1.4\mu M$ by  $\Delta S=0.2\mu M$.
As for the spin  of a particle with mass
$\mu$, we usually expect that $S \mbox{$\;\lower-.2ex\hbox{$\textstyle<$}\;\!\!\!\!\!\!
\lower.7ex\hbox{$\textstyle \sim$}\;$} O(\mu^2)$. In fact, for
a rotating Kerr black hole with mass $M$ and angular
momentum $J$, we have the inequality $J \leq M^2$, where the equality holds in
the extreme limit.  Therefore, for the present case, $S/\mu M =
(S/\mu^2)\cdot \mu /M \mbox{$\;\lower-.2ex\hbox{$\textstyle<$}\;\!\!\!\!\!\!
\lower.7ex\hbox{$\textstyle \sim$}\;$} O(\mu /M)$ should be much smaller than unity for a
test particle, for which we assume that $\mu \ll M$.  However, if
we would extend our analysis to the case of a relativistic binary 
system with masses $m_1$ and $m_2$, we may find that $S/\mu M \sim O(1)$ as follows. 
$\mu$ and $M$ should be
regarded as the reduced and total masses, i.e., $\mu =m_1 m_2
/(m_1+m_2), M=m_1+m_2$, and for the spin of the first
relativistic star 1, $S_1/ \mu M = S_1 / m_1m_2 = (S_1 /
m_1^2) \cdot m_1/m_2$. Then we have $S_1/ \mu M = S_1 /
m_1^2 \mbox{$\;\lower-.2ex\hbox{$\textstyle<$}\;\!\!\!\!\!\!
\lower.7ex\hbox{$\textstyle \sim$}\;$} O(1)$ for the case of $m_1=m_2$.  Therefore $S_1/ \mu M $ can be large as
unity. Because the value of $S=1\mu M$ is not a mathematical
special bound\cite{comment1}, we will analyze the case of $S>1\mu M$ as well in order to
see the spin effect more clearly. 

   From Figs.4a and 4b, which correspond to  the type (B1)
potential, we find that no tori are broken. Chaos
never occurs. On the other hand, in the case of the type (B2)
potential, which is shown in Figs.4c-4f, we  see that some tori are  broken, which means that chaos occurs for such
orbits. This is the first example of chaotic behavior in the
motion of a test particle in a Schwarzschild spacetime. The sea of
chaos spreads in these Poincar\'e sections as $S$ increases. We conclude that this chaotic behavior is induced by the spin effect.

Fig.5 shows the orbit in the 2-dimensional configuration space  corresponding  to each torus ((1) $\sim$ (3)) in Fig.4d ($S=1 \mu M$).
The orbit  (1)
is almost perpendicular  to the equatorial plane. The
initial position is near the minimum point of the ``effective
potential".  The orbit (2)  is chaotic and  the particle
approaches the saddle points. 
The orbit (3)  is constrained in the very narrow
area near the equatorial plane contrary to the
orbit (1).  The initial positions are located
near the edge of the ``effective potential". 
In Fig.6, we present the Lyapunov exponents $\lambda$ for  orbits (1) $\sim $ (3) in Fig.5.  $\tau_\lambda / T_{\rm P}$ is also
given, where $T_{\rm P}$ is the mean orbital
period averaged after $10^2$ rotations around a black hole. 
$\tau_\lambda / T_{\rm P}$ gives a naive estimation for how many
rotations we expect before the chaotic behavior becomes
distinct.
 We may justify that the orbit (2) is
strongly chaotic because of the large positive value of its Lyapunov exponent. The reason why the Lyapunov exponents of the orbits (1) and (3) converge to  positive values may be either that
 the time series we prepared to calculate the Lyapunov exponent
are not sufficient or that the system is not integrable even
through we cannot see any break down of the tori in the Poincar\'e map. 

Is there any critical value of the spin for occurrence of chaos? 
If it is determined by the potential type, we obtain
$S_{cr}=0.64 \mu M$ for $J=4 \mu M$ from Fig.3.  To confirm this, we plot the
Lyapunov exponent $\lambda$ in terms of the value of the spin $S$
in Fig.7.  After $S$ reaches the critical value $S_{cr}$,
$\lambda$  increases remarkably, which supports our conclusion.

From our numerical investigation,  we find several
 conditions  for the occurrence of chaos when setting up the initial data. 
\begin{itemize}
\item A particle must move in the type (B2) potential.
\item The particle must be bounded in a compact region, i.e., $E < E_{\mbox{sp}}$, where
$E_{\mbox{sp}} = V({\bf r}_{\mbox{sp}})$ is the potential energy
at the saddle point ${\bf r}_{\mbox{sp}}$.
\item The particle must have energy $E\sim E_{\mbox{sp}}$ and the position $r_{0}$, and the angle of the spin 
$\alpha$ must be appropriately chosen in order for the particle to approach the saddle point. 
\end{itemize}

%%%%%%%%%%%%%%%%%%%%%%%%%%%%%%%%%%%%%%%%%%%%%%%%%%%%%%
%                                                    %
%    Section 3-2 (Heteroclinic Orbit)                %
%                                                    %
%          Fig.8:Extreme orbit                       %
%                                                    %
%%%%%%%%%%%%%%%%%%%%%%%%%%%%%%%%%%%%%%%%%%%%%%%%%%%%%%
\subsection{Transition from Regular to Chaos}
Here we shall briefly 
discuss a mechanism for the occurrence of chaos. 

Fig.7 shows the Poincar\'e map of five orbits with
$J=3.81\mu M$, $S=1\mu M$ and $E=0.923\mu$.
Note that $E_{\mbox{sp}}=0.9288\mu$ for this case. 
We  see four different types of orbits. 
\begin{itemize}
\item Type (i) :
The stable orbits which form a closed curve near the center of the Poincar\'e map.
\item Type (ii) :
The stable orbits which form a closed curve at the edge of the
Poincar\'e map. 
\item Type (iii) :
The stable orbits which form a closed curve in the upper or
lower part of the Poincar\'e map. 
\item Type (iv) :
The separatrix of the above three types of orbits. Two hyperbolic
fixed points in this Poincar\'e section exist.
\end{itemize}
It is plausible that the type (iv) orbit plays an important role
for the occurrence of chaos. The unstable manifold
departing from one hyperbolic fixed point will join smoothly to
the stable manifold of another hyperbolic fixed point. Such
an orbit is called a ``heteroclinic orbit''.

As $E$ increases from this state,  the sea of chaos appears around
the hyperbolic fixed points and spreads into whole
Poincar\'e map.
Its behavior is shown in Fig.8. The chaos observed here may be caused by a ``heteroclinic tangle'',
although we need further detailed analysis. 

The next question is what causes the  existence
of type (iv) orbits. Although we do not have a definite answer now, we
find some correlation between an appearance of such an orbit and the type of
``effective potential".
As $S$ decreases, the type (ii), and then types  (iii) and (iv)
will vanish and  only  the type (i) orbit remains. This transition
is seen in Fig.4. For the case of  $S=0.8\mu M$ (Fig.4c), we
cannot find the type (ii) orbit, and in the cases of $S=0.6$ and
$0.4\mu M$ (Figs.4a and 4b), all orbits belong to the type (i). 
The latter case belongs to  the type (B1)
potential.
Hence, it seems that there exists some relation between the
type of ``effective potential" and an appearance of types (ii),
(iii) and (iv) orbits.  We may find some criterion for chaos
using the ``effective potential".

%%%%%%%%%%%%%%%%%%%%%%%%%%%%%%%%% 
%                                  %
%        Chapter 4              % 
%      Concluding Remarks       % 
%                                  %
%%%%%%%%%%%%%%%%%%%%%%%%%%%%%%%%%
\section{Summary and Discussion}
\label{sec5}\setcounter{equation}{0}
In this paper, using the pole-dipole
approximation,  we study the motion of a spinning test
particle near a Schwarzschild black hole to clarify its
dynamical properties such as a chaos. We find the motion of the
particle can be chaotic under some appropriate conditions.
 Because the motion
of a spinless particle in this spacetime is never  chaotic
because of its integrability, 
this chaotic behavior is purely induced by the
spin-orbit interaction. The ``effective potential" of the particle
is also introduced to classify the dynamical behaviors. 
The ``effective potential"s are classified into  four 
different types depending on the total angular momentum  $J$ and
the spin $S$.
When $J$ is large, some orbits are bounded and the ``effective
potential"s are classified into two types: for type (B1) one saddle
point (unstable circular orbit) and one minimal point (stable
circular orbit) on the equatorial plane exist for small spin;  and for (B2) two saddle points bifurcate from the equatorial plane and one minimal point remains on the
equatorial plane for large spin. If $J$ is small, no
bound orbits exist and the potentials are classified into another
two types: for type (U1) no extremal point is found for small spin; and   for type (U2)  one saddle point appears on the equatorial plane, which is unstable
in the direction perpendicular to the equatorial plane, for large
spin. The types (B1) and (U1) are the same as those for a
spinless particle, but the types (B2) and (U2) are new potentials which appear through a spin-orbit coupling.  The chaotic behavior 
is found only in the type (B2) potential.
We believe that the appearance of saddle points is important, because we find chaos
only for the orbits which 
approach the saddle points.

The critical value of the spin beyond which chaos will occur
is $S \sim 0.64 \mu M$ for $J= 4 \mu M$.
We also present the Lyapunov exponent, which increases
rapidly after the spin $S$ gets larger than the above critical
value.  This supports the use of the ``effective potential'' as a criterion for chaos. The typical value of the Lyapunov
exponent is about several orbital periods of the particle, which
may become important in some relativistic astrophysical phenomena.

In a real astrophysical system such as  a binary system, the symmetry of the system is  lower than the present case. There may
be  other important effects in addition to the  spin-orbit interaction, 
which make the motion more complicated. Then we may expect that
chaos occurs even in the real system, or that the other effects 
stabilize the system and chaos will never be found. 
We need further analysis taking into account the other effects
such as the spin-spin interaction or a force due to  multipole
moments, even if we adopt a test particle 
analysis\cite{comment2}.
As for the spin-spin interaction, we can analyze a spinning test
particle in  a rotating Kerr black hole. In our preliminary
analysis by the ``effective potential" near the equatorial plane, the
critical value of the spin for chaos gets smaller as the angular
momentum of the black hole becomes larger.  The detailed analysis is
under investigation.

Another important point is whether or not such a chaotic behavior, if it exists, 
affects any realistic astrophysical or physical phenomena.
One of the important targets for such an investigation is a coalescing binary system, where we need
general relativity, in particular relativistic dynamics of
compact objects as we mentioned in section 1.
To examine it, since we have studied here only the condition
for occurrence of chaos, then we next have to know the evolution of the
system including emission of gravitational waves.  The particle
traveling around a black hole emits the gravitational waves,
extracting  the energy and the angular momentum from the system. This will tell us whether or not  the evolutionary path will get into the region of parameter space where chaos will take place.
We then have to calculate the emission rates of  
the energy and the angular momentum, $\dot{E}$ and
$\dot{J}$ for the present system\cite{mino},\cite{mino2} and follow the evolution.

\vskip 1cm
We would like to thank M. Sasaki, M. Shibata, Y. Sota, H. Tagoshi and T. Tanaka for useful discussions and thank P. Haines for reading the paper carefully. 
SS also acknowledges K. Imafuku a stimulating discussion.
This work was supported
partially by the Grant-in-Aid for Scientific Research  Fund
of the Ministry of Education, Science and Culture  (No.
06302021 and No. 06640412),   and by the Waseda University
Grant for Special Research Projects. 

%%%%%%%%%%%%%%%%%%%%%%%%%
%                                                    %
%                   Appendix A               %
%                                                    %
%%%%%%%%%%%%%%%%%%%%%%%%%
\newpage
\renewcommand{\theequation}{\mbox{$A .\arabic{equation}$}}
\setcounter{equation}{0}
\begin{flushleft}
{\bf Appendix A: The Basic Equations by Use of Spin Vector}
\end{flushleft}
\vspace{.5 cm}

In the text, we have used  a spin tensor $S^{\mu\nu}$. 
However, it may be sometimes more convenient or more 
intuitive to describe the basic
equations by use of a spin vector $S_\mu$, which is defined by
\begin{equation}
S_\mu = - \frac{1}{2}\epsilon_{\mu\nu\rho\sigma}
u^{\nu}S^{\rho\sigma}, 
\label{def:spin}
\end{equation}
which gives the following constraint:
\begin{equation}
p^\mu S_\mu = 0. 
\end{equation}

The equations of motion
(\ref{eqn:xdot}), (\ref{eqn:pdot}), and (\ref{eqn:sdot})
are now 
\begin{eqnarray}
\frac{dx^{\mu}}{d\tau}&=&v^{\mu},
\label{eqn:xdot2}
\\
\frac{Du^{\mu}}{D\tau}&=&
-\frac{1}{\mu} 
R^{*\mu}_{~~\nu\rho\sigma}v^{\nu}u^{\rho}S^{\sigma},
\label{eqn:pdot2} \\
\frac{DS^{\mu}}{D\tau}&=&-\frac{1}{\mu} u^{\mu}
(R^{*}_{~\alpha\beta\gamma\delta}S^\alpha
v^{\beta}u^{\gamma}S^{\delta}), 
\label{eqn:sdot2}, 
\end{eqnarray}
where
 \begin{equation}
R^{*\alpha\beta}_{~~~\gamma\delta} \equiv 
\frac{1}{2}
R^{\alpha\beta\rho\sigma} 
\epsilon_{\rho\sigma\gamma\delta}.
\end{equation}
The relation between the 4-velocity and the momentum
is 
 \begin{equation}
v^{\mu} = N \left(u^{\mu} - \frac{1}{\mu^2}
~^{*}R^{*\mu\alpha}_{~~~\beta\gamma} S_\alpha u^\beta S^\gamma
\right),
\label{eqn:v-u2}
\end{equation}
where 
\begin{equation}
 ~^{*}R^{*}_{~\mu\nu\rho\sigma} \equiv
\frac{1}{2} \epsilon_{\mu\nu\alpha\beta}
R^{*\alpha\beta}_{~~~\rho\sigma} = 
\frac{1}{4} \epsilon_{\mu\nu\alpha\beta}
R^{\alpha\beta\gamma\delta}
\epsilon_{\gamma\delta\rho\sigma}.
\end{equation}
and $N$ is the normalization constant determined from $v_\mu v^\mu =-1$.
The supplementary condition to fix the center of mass is 
\begin{equation}
v_\mu S^\mu =0.
\end{equation}
The spin vector is perpendicular to the 4-velocity as we expected.
Note that $v_\mu S^{\mu\nu} \neq 0$.
Equation (\ref{eqn:spin}) is just 
\begin{equation}
S^2 = S^\mu S_\mu .
\label{eqn:spin2} 
\end{equation}

In what follow, we assume a Schwarzschild black hole as the background spacetime.
We write down explicitly  the relation (\ref{def:spin}) between  $S^{\mu\nu}$ and  $S_{\mu}$ as 
\begin{eqnarray}
S_t &= &-r^2 \sin \theta \left[u^r S^{\theta\phi}
+u^\theta S^{\phi r} + u^\phi S^{r \theta}\right]
\label{eqn:sst}\\
S_r &= &r^2 \sin \theta \left[u^t S^{\theta\phi}
-u^\theta S^{t \phi} + u^\phi S^{t \theta}\right]
\label{eqn:ssr}\\
S_\theta &= & r^2 \sin \theta  \left[u^t S^{\phi r}
-u^\phi S^{t r} + u^r S^{t \phi}\right]
\label{eqn:ssth}\\
S_\phi &= & r^2 \sin \theta  \left[u^t S^{r \theta}
-u^r S^{t \theta} + u^\theta S^{t r}\right]
\label{eqn:ssph}
\end{eqnarray}
Conversely, Eqs. (\ref{eqn:sst}) $\sim$ (\ref{eqn:ssph}) with Eqs.(\ref{eqn:ps0}), (\ref{eqn:sst})  give 
\begin{eqnarray}
S^{\theta\phi}&= &
-\frac{u_t}{r^2 \sin\theta}\tilde{S}_r
\label{eqn:ssr3}\\
S^{\phi r} &= & 
-\frac{u_t}{r^2 \sin\theta}\tilde{S}_\theta
\label{eqn:ssth3}\\
S^{r \theta} &= & 
-\frac{u_t}{r^2 \sin\theta}\tilde{S}_\phi ,
\label{eqn:ssph3}
\end{eqnarray}
where 
\begin{equation}
\tilde{S}_\mu \equiv S_\mu -\frac{S_t}{u_t}u_\mu .
\end{equation}

Using Eqs.(\ref{eqn:jx2})$\sim$(\ref{eqn:stph}), 
 Eqs.(\ref{eqn:sst})$\sim$(\ref{eqn:ssph}) are now
\begin{eqnarray}
\tilde{S}_r &=
&\frac{J}{u_t}\left[-\left(1+fu_r^2\right)\cos\theta 
\right]
\label{eqn:ssr2}\\
\tilde{S}_\theta &=
&\frac{r}{u_t}\left[J\left(
\sin\theta - fu_r \frac{u_\theta}{r}\cos\theta\right)
-\frac{p_\phi}{\sin\theta}\right]
\label{eqn:ssth2}\\
\tilde{S}_\phi &= &\frac{r}{u_t}\left(p_\theta\sin\theta
-J f u_r\frac{u_\phi}{r}
  \cos\theta\right)
 \label{eqn:ssph2}\\
S_t &= &J\left(-f u_r \cos\theta
+\frac{u_\theta}{r}\sin\theta\right)
\label{eqn:sst2}
\end{eqnarray}
This expression already includes the constants of motion for the angular momentum.

The initial direction of the spin vector at $r=r_0,\;\theta =\pi/2,\;\phi=0$ is given by the equations
\begin{eqnarray}
\tilde{S}_r&= & 0
\label{eqn:sr3}
\\
\tilde{S}_\theta &=
&\frac{r}{u_t}\left( J -p_\phi\right)
\label{eqn:sth3}
\\
 \tilde{S}_\phi &= &\frac{r}{u_t}p_\theta .
\label{eqn:sph3}
\end{eqnarray}

As for the angle of the spin $\alpha$, we find that
\begin{equation}
\alpha = \tan^{-1} \frac{ \tilde{S}_\phi}{\tilde{S}_\theta} . \end{equation}
Using this definition with (\ref{eqn:spin2}), we have 
\begin{eqnarray}
\tilde{S}_\theta &=
&\frac{\mu Sr_0^2}{\sqrt{\mu^2 r_0^2 + J^2 
\sin^2\alpha}}\cos\alpha 
\\
 \tilde{S}_\phi &= &\frac{\mu Sr_0^2}{\sqrt{\mu^2 r_0^2 + J^2 
\sin^2\alpha}}\sin\alpha .
\end{eqnarray}
$u_\theta, u_\phi$ are then some functions of $u_t$ through Eqs. (\ref{eqn:sth3}) and (\ref{eqn:sph3}).  Inserting them into the energy conservation equation (\ref{eqn:energy2}), we obtain the quadratic equation for  $u_t$.  We also have Eq. (\ref{eqn:mass}) for $u_r$.  Solving them, we can finally set up the initial data of $u^\mu$ and $S^\mu$.

Setting $p_r=p_\theta=0$ and using Eqs. (\ref{eqn:mass}) and (\ref{eqn:spin2}), we find the ``effective potential"
(\ref{eqn:effpot}).
After giving $J,S$ and $E$ and setting initial data, we can solve the dynamical equations  (\ref{eqn:xdot2}) and  (\ref{eqn:pdot2}) with algebraic equations (\ref{eqn:v-u2}), (\ref{eqn:ssr2}) $\sim$ (\ref{eqn:sst2}), although here we have solved for all variables using the dynamical equations in order to estimate the accuracy by the constraint equations.

%%%%%%%%% Refereces %%%%%%%%%%%
.

\newpage
\begin{flushleft}
{ Figure Captions}
\end{flushleft}

\vskip 0.1cm
   \noindent
%%%%%%
\parbox[t]{2cm}{ FIG. 1:\\~}\ \
\parbox[t]{14cm}
{The ``effective potential" $V_{(\pm)}$
on the equatorial plane for $J=4 \mu M$ and $S=1 \mu M$. We see
that the particle in $V_{(-)}$ moves in the region closer to the
event horizon ($r=2  M $) than that in $V_{(+)}$.
}\\[1em]
\noindent
%%%%%%
\parbox[t]{2cm}{ FIG. 2:\\~}\ \
\parbox[t]{14cm}
{Four different types of the ``effective potential" $V$. The
saddle point of the potential is marked by a cross. (a) Only
one saddle point exists on the equatorial plane. This potential is
 similar to that for a spinless particle. Chaos never occur in
this case. (b) Two saddle points are found in  both sides of the
equatorial plane. The orbit can be chaotic. (c) When $J$ is very
small, the centrifugal force is too small
to balance with the gravity and then no bound region is found.
This potential is
also  similar to that for a spinless particle. 
(d) This also has no bound region. But its shape is
 different from  (c). There exists a saddle
point on the equatorial plane. This point is locally minimal in the $r$ direction but maximal in $\theta$ direction. The
particle will eventually fall into the black hole after leaving
the equatorial plane.
}\\[1em]
\noindent
%%%%%%
\parbox[t]{2cm}{ FIG. 3:\\~}\ \
\parbox[t]{14cm}
{The types of the ``effective potential" are classified by $J$
and $S$. The value of $S$ at the bottom
end of the type (U2) region is slightly smaller than $1 \mu M$.
}\\[1em]
\noindent
%%%%%%
\parbox[t]{2cm}{ FIG. 4:\\~}\ \
\parbox[t]{14cm}
{The Poincar\'e Maps for various values of $S$. All orbits
have the total angular momentum $J=4.0\mu M$. 
We set $p^{r}=0$ initially. (a) $S=0.4 \mu M$ and $E=0.97698396\mu $: The initial position for each torus is $r_0=3.8,\;6.0$ and $8.0M$. (b) $S=0.6 \mu M$ and $E=0.96730999\mu $ and the initial position is 
 $r_0=3.6,\;5.5,\;8.0$ and $12.0M$. (c) $S=0.8 \mu M$, 
$E=0.95815568\mu $ and
$r_0=3.7,\;4.2,\;6.0$ and $8.0M$. (d) $S=1.0 \mu M$,
$E=0.94738162\mu $ and 
$(1)\;\;r_0=5M,\;(2)\;\;3.9M$(chaotic)$,\;(3)\;\;3.72M$. 
The orbits in 2-dimensional configuration space are shown Fig.5. 
(e) $S=1.2 \mu M$, $E=0.93545565\mu$ and
$r_0=3.86,\;4.2,\;5.2$ and $6.0M$. (f) $S=1.4 \mu M$, $E=0.92292941\mu$
and $r_0=4.5,\;5.0,\;5.7$ and $7.6M$. Very strong chaos occurs
in  this case, although it may be unrealistic.
}\\[1em]
\noindent
%%%%%%
\parbox[t]{2cm}{ FIG. 5:\\~}\ \
\parbox[t]{14cm}
{The orbit in the 2-dimensional configuration space 
corresponding  to each torus in Fig.4d ((1) $\sim$ (3)).
 (1) This orbit is almost perpendicular  to the equatorial plane. The
initial position is near the minimum point of the ``effective
potential". 
(2) The chaotic orbit. The particle
approaches the saddle points marked by a cross. 
(3) Contrary to the orbit shown in (1), this orbit is constrained in the very narrow area near the equatorial plane. The initial position is located
near the edge of the ``effective potential".
}\\[1em]
\noindent
%%%%%%
\parbox[t]{2cm}{ FIG. 6:\\~}\ \
\parbox[t]{14cm}
{The Lyapunov exponent of each orbit shown in Fig.5. We find that the orbit (2) is strongly chaotic. The ratio of the inverse Lyapunov exponent $1/\lambda$ to the average orbital period $\tau_{\lambda}\equiv T_{P}$ is also shown.
}\\[1em]
\noindent
%%%%%%
\parbox[t]{2cm}{ FIG. 7:\\~}\ \
\parbox[t]{14cm}
{The Lyapunov exponent $\lambda$  in terms of the value of spin
$S$.  When $S$ becomes larger than  the critical value $S_{cr} \sim
0.64 \mu M$, beyond which we find chaos, $\lambda$ increases
rapidly.  This supports the notion of a critical value of
the spin for occurrence of chaos.
}\\[1em]
\noindent
%%%%%%
\parbox[t]{2cm}{ FIG. 8:\\~}\ \
\parbox[t]{14cm}
{The classification of the orbits for $J=3.81 \mu M,\;S=1 \mu M$ and $E=0923 \mu$.  We find two separatrix (Type (iv)) which divide
three types of stable orbits (Type (i) $\sim$ (iii)).  
There exists a heteroclinic orbit starting from one hyperbolic
fixed point to the other fixed point, which may cause the present
chaotic behaviors.
}\\[1em]
\noindent
%%%%%%
\parbox[t]{2cm}{ FIG. 9:\\~}\ \
\parbox[t]{14cm}
{When the energy increases from that in Fig.8, the torus
gradually spreads out around the heteroclinic orbit in the
Poincar\'e map. 
}\\[1em]
\noindent
%%%%%%


\begin{thebibliography}{99}
\bibitem{comment1} To find a non-singular ``effective potential'', $S$ should be smaller than $2\sqrt{2}\mu M$.
\end{thebibliography}
\end{document}